# All-electrical spin-to-charge conversion in sputtered $Bi_xSe_{1-x}$

Won Young Choi[1†*], Isabel C. Arango[1†], Van Tuong Pham[2], Diogo C. Vaz[1], Haozhe Yang[1], Inge Groen[1], Chia-Ching Lin[3], Emily S. Kabir[3], Kaan Oguz[3], Punyashloka Debashis[3], John J. Plombon[3], Hai Li[3], Dmitri E. Nikonov[3], Andrey Chuvilin[1,4], Luis E. Hueso[1,4], Ian A. Young[3] and Fèlix Casanova[1,4*]

[1]CIC nanoGUNE BRTA, 20018 Donostia-San Sebastián, Basque Country, Spain.

[2]Univ. Grenoble Alpes, CNRS, Institut Neel, F-38000 Grenoble, France.

[3]Components Research, Intel Corp., Hillsboro, OR 97124, USA.

[4]IKERBASQUE, Basque Foundation for Science, 48009 Bilbao, Basque Country, Spain.

[†]These authors contributed equally: Won Young Choi, Isabel C. Arango.

*e-mail: w.choi@nanogune.eu, f.casanova@nanogune.eu

## ABSTRACT

One of the major obstacles to realizing spintronic devices such as MESO logic devices is the small signal magnitude used for magnetization readout, making it important to find materials with high spin-to-charge conversion efficiency. Although intermixing at the junction of two materials is a widely occurring phenomenon, its influence on material characterization and the estimation of spin-to-charge conversion efficiencies is easily neglected or underestimated. Here, we demonstrate all electrical spin-to-charge conversion in $Bi_xSe_{1-x}$ nanodevices and show how the conversion efficiency can be overestimated by tens of times depending on the adjacent metal used as a contact. We attribute this to the intermixing-induced compositional change and the properties of a polycrystal that lead to drastic changes in resistivity and spin Hall angle. Strategies to improve the spin-to-charge conversion signal in similar structures for functional devices are discussed.

The improvement of computational performance over the past few decades has relied on an increase in the number of transistors led by the successful miniaturization of complementary metal–oxide–semiconductor (CMOS) transistors[1]. However, further improvement in performance is limited by unscaled power density[2,3], triggering an increasing demand for energy efficient beyond-CMOS devices that address this problem. The magnetoelectric spin-orbit (MESO) logic device[4–6] proposed in 2018 as an alternative to CMOS device operates at very low power because it uses a magnetoelectric material for magnetization switching (writing). On the other hand, the reading part of the device[7] includes a spin-orbit material that requires a high output voltage of 100 mV[4–6] for full operation. In order to obtain such a large output voltage through spin-to-charge conversion (SCC), high resistivity as well as high conversion efficiency are required.

Topological insulators (TI) are a new class of materials[8,9] that have topologically protected surface states with spin-momentum locking, a property that should lead to very efficient SCC[10]. Indeed, the TI is considered to have an exceptionally large spin Hall angle, $\theta_{SH}$, according to spin-orbit torque results[11–15] and long inverse Rashba-Edelstein length, $\lambda_{IREE}$, in proximitized system[16].Together with high resistivity, as bulk conduction is limited[17,18], they satisfy the required conditions for MESO. Simultaneously, they have limitations in stoichiometric and single crystalline material growth and require low operating temperature for electrical spin injection[19,20]. Unlike epitaxially grown TI, sputtered $Bi_xSe_{1-x}$ was reported to have high conversion efficiency and resistivity even at room temperature, despite its simple growth technique and polycrystalline structure[21]. However, recent reports on the formation of an interfacial layer between $Bi_2Se_3$ and transition metals by a solid-state reaction[22–25] encourages rigorous characterization of sputtered $Bi_xSe_{1-x}$.

Here, we show the influence of the intermixing at the junction of $Bi_xSe_{1-x}$ (from now on, BiSe) and transition metals on the characterization of the electrical and spintronic properties of BiSe using a local spin injection device. Sputtered BiSe was characterized by transmission electron microscopy (TEM), energy-dispersive X-ray spectroscopy (EDX), electrical measurements, and 3D finite element method (FEM) simulations. We observed a huge variation in resistivity and $\theta_{SH}$, according to the compositional change of BiSe due to adjacent metal layer, through the BiSe thickness dependence of the spin-to-charge conversion signal and the cross-junction resistance. This result emphasizes the importance of proper material characterization, as it can greatly affect the SCC efficiency evaluation, particularly in the case of structures or devices including reactive materials such as BiSe and transition metals.

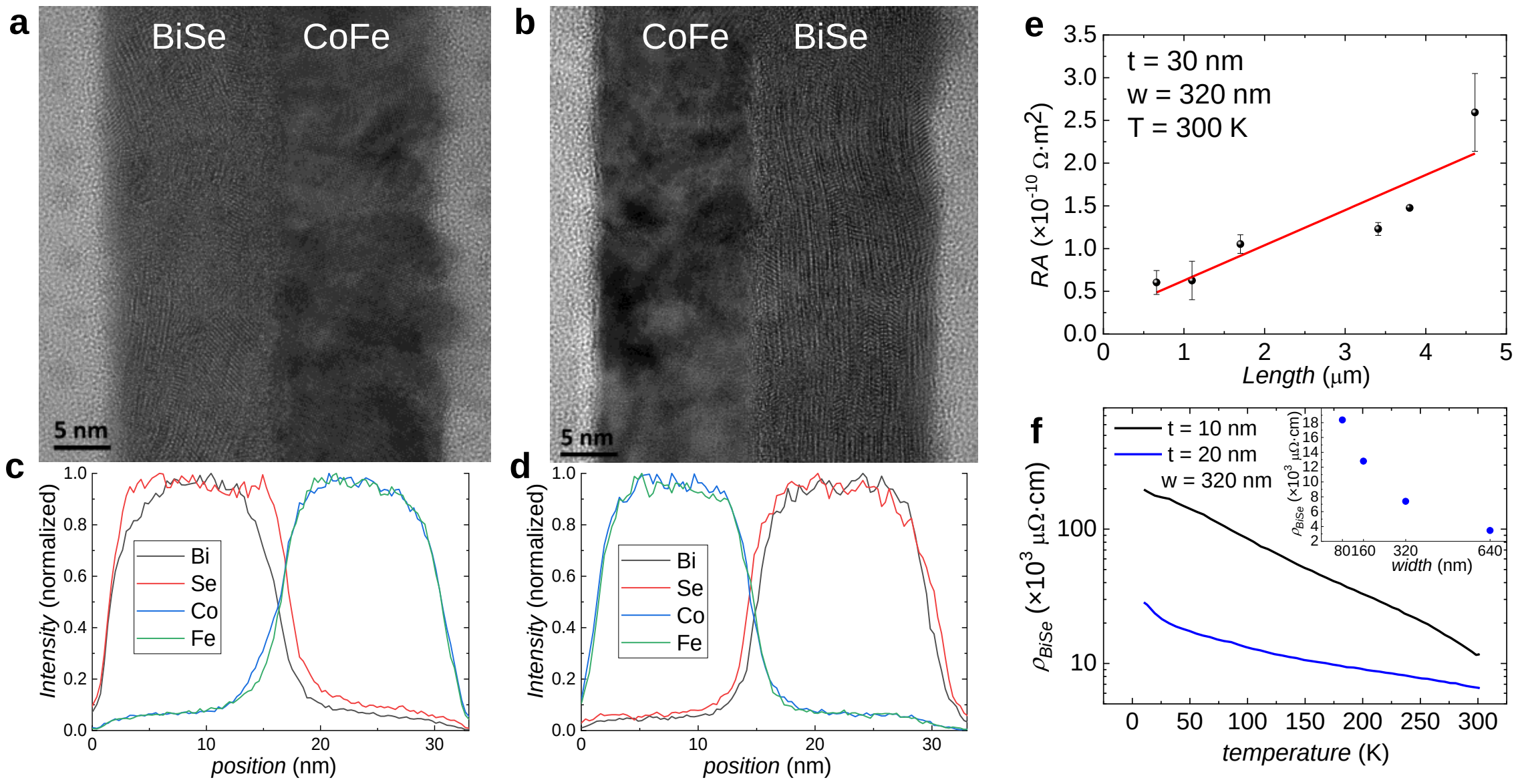

**Figure 1.** Characterization of sputtered BiSe. (a–d), TEM images and normalized EDX results of BiSe/CoFe (a–c) and CoFe/BiSe (b, d) bilayer stacks. (e), Length dependence of the resistance area product measured for a 320-nm-wide and 30-nm-thick wire at room temperature. From the linear fitting, a resistivity of 4000 μΩ·cm is determined. (f), Resistivity of BiSe as a function of temperature for 320-nm-wide wires with a thickness of 10 nm (black) and 20 nm (blue). Inset: Width dependence of the resistivity of a 20-nm-thick BiSe wire.

Thin film bilayers composed of BiSe (16 nm thick) and CoFe (15 nm thick) were prepared to characterize sputtered BiSe prior to SCC experiments. TEM images in Figure 1a and 1b show the structure of $SiO_2$ substrate/BiSe/CoFe/capping layer and $SiO_2$ substrate/CoFe/BiSe/capping layer, respectively. The sputtered BiSe layer is polycrystalline in both cases, in agreement with the literature[21], although it is more oriented on top of the CoFe layer than on top of the $SiO_2$ substrate. In both cases, an amorphous layer was found at the interface, being thicker in the BiSe/CoFe stack. Such layer corresponds to intermixing at the interface, as confirmed by normalized Energy-dispersive X-ray spectroscopy (EDX), which shows a clear shift of the Se curve (red) in both cases as shown in Figure 1c and 1d, being larger in the BiSe/CoFe stack in agreement with the thicker amorphous layer. Since such an imperfection at the interface will adversely affect the spin injection between the two materials and the subsequent SCC, we chose the CoFe/BiSe stack (Figure 1b and 1d).

The high resistivity of sputtered BiSe is one of the reasons why it is considered the strongest candidate for the magnetic readout part of MESO logic devices[4–6]. Figure 1e shows the resistance area product of a 30-nm-thick and 320-nm-wide BiSe wire measured at room temperature at different distances using metallic contacts. The resistivity of the BiSe nanowire (4,000±1000 μΩ·cm) was obtained from a linear fitting (Note S1in the Supporting Information). Unlike metallic conductors, BiSe has a high noise level in electrical measurements and a big dispersion from the fitting line. As shown in Figure 1f, the resistivity increases as the temperature decreases, indicating a semiconducting behavior, and varies with the thickness (10 and 20 nm) and width (80–640 nm). The BiSe wire with 80 nm width and 20 nm thickness has a resistivity of 18,000 μΩ·cm at room temperature, which shows that the sputtered BiSe used in this experiment is similar to those previously reported[21].

To study SCC in BiSe, we use the local spin injection device, which corresponds to the architecture of the spin-orbit reading module of the MESO device. All materials constituting the local spin injection device were grown by dc (metallic layers) and rf (BiSe and $SiO_2$ capping layers) sputter deposition. The device consists of a top T-shaped nanostructure of BiSe (2 to 40 nm)/NM (Ti, Pt or Ta; 10 nm) and a bottom15-nm-thick CoFe electrode, all with a width of 80 nm. Since sputtered BiSe has a high noise level in electrical measurements, a normal metal (NM) such as Ti, Pt, and Ta is deposited on top of BiSe to pick up the SCC output voltage through electrical shunting. All measurements presented below are obtained at room temperature. Fabrication steps, characterization techniques, and measurement details are described in Methods.

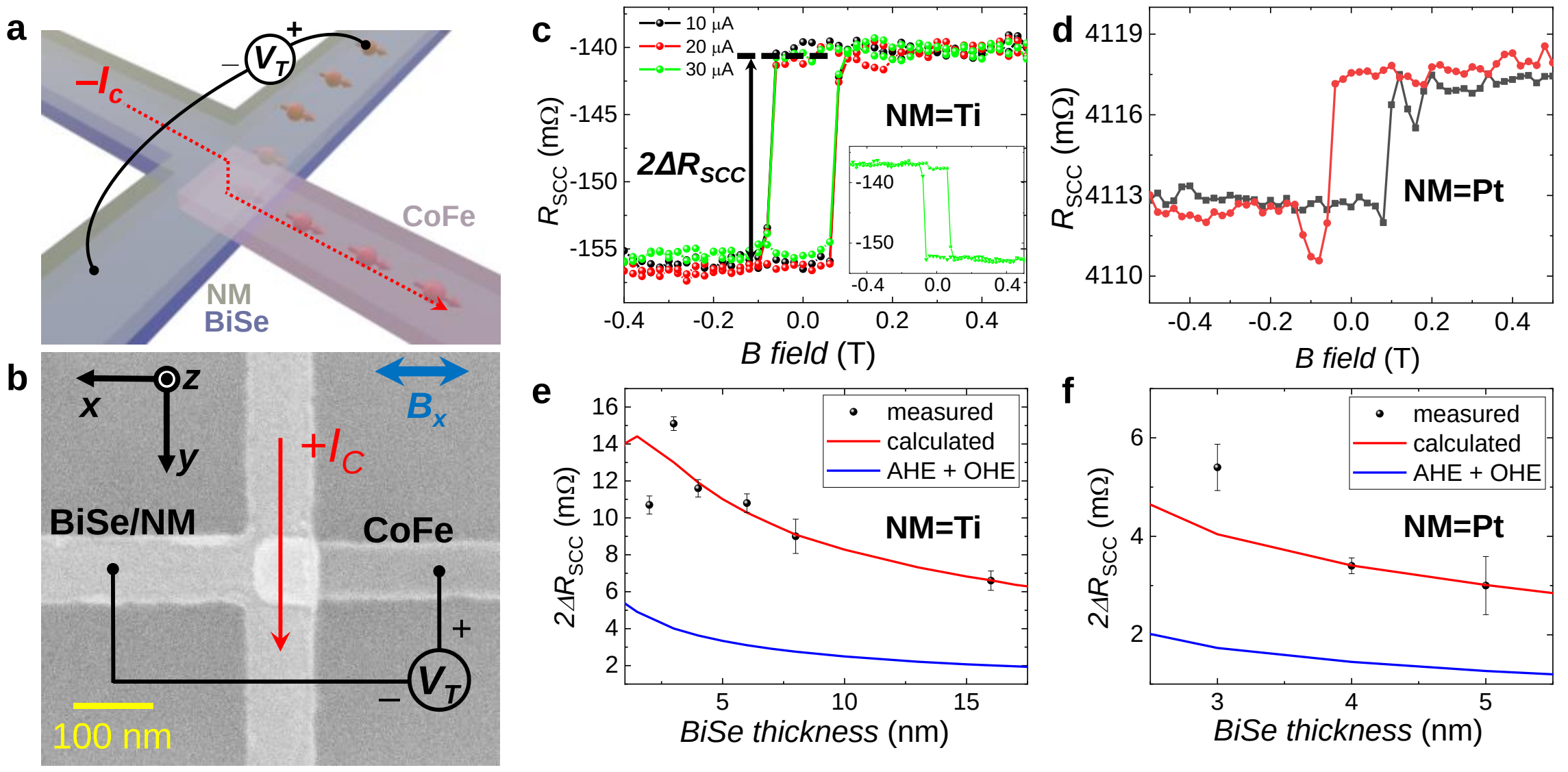

**Figure 2.** Local spin injection device and spin-to-charge conversion signals. (a) Schematics of local spin injection device and measurement configuration of spin-to-charge conversion. Vertical spin current at the junction produces charge current in the transverse direction. (b) SEM image of the device with measurement geometry of charge-to-spin conversion. (c, d) SCC resistance ($R_{SCC}$) measured using the configuration in panel (a) as a function of magnetic field along $x$-axis of (c) BiSe(3 nm)/Ti(10 nm) and (d) BiSe(3 nm)/Pt(10 nm) structures. Inset: charge-to-spin conversion resistance measured using the configuration in panel (b) as a function of magnetic field along $x$-axis. (e, f) The SCC signal ($2\Delta R_{SCC}$) as a function of BiSe thickness and fitting curves obtained by 3D FEM simulation of (e) BiSe/Ti and (f) BiSe/Pt structures. All measurements are performed at room temperature.

To quantify the SCC signal, the transverse voltage ($V_T$) is measured while an external magnetic field is applied along the easy axis of the CoFe electrode ($x$-axis) and a current ($I_C$) flows from CoFe to one end of the BiSe/NM T-shaped nanostructure as shown in Figure 2a. The reciprocal measurement (charge-to-spin conversion, CSC) is described in Figure 2b. Spin-polarized current in the CoFe wire flows vertically through BiSe to the NM due to the large difference in resistivity, and SCC occurs inside BiSe. The resulting charge current is then measured transversely in the T-shaped structure as an open circuit voltage $V_T$. When the magnetization of the CoFe electrode is switched, the spin polarization also reverses and $V_T$ changes sign. The SCC resistance, defined as $R_{SCC} = V_T/I_C$, always contains a baseline value and therefore it is more convenient to define the SCC signal, $2\Delta R_{SCC}$, as the difference between the two magnetic states (see Figure 2c). The converted current is mostly shunted by CoFe, since the NM completely covers the BiSe T-shaped nanostructure, and partially by the NM greatly reducing the magnitude of $V_T$. What we finally measure is the voltage across the Hall cross shunted by the NM, being thus dependent on the resistivity of the NM. Nevertheless, the use of a NM layer is crucial, as it dramatically lowers the noise level and makes the SCC signal measurable, in contrast to the use of lateral NM contacts in a BiSe-only T-shaped nanostructure.

Figure 2c shows the SCC signal of 15.1 ± 0.4 mΩ measured on a BiSe(3 nm)/Ti(10 nm) device. The hysteresis loop was observed according to the switching field of CoFe wire, and the current dependence and reciprocal measurement (CSC signal shown in the inset of Figure 2c) confirm that they were conducted in the linear response regime, which rules out the presence of any heating-related effects. The same measurement on a BiSe(3 nm)/Pt(10 nm) device is shown in Figure 2d, with an SCC signal of 5.4 ± 0.5 mΩ, which is 3 times smaller than that of a BiSe/Ti device with the same thickness.

Since the spin Hall effect can occur in the adjacent NM when BiSe is thin, we need to check whether the SCC signal is generated in BiSe. Comparing the SCC signal of the two devices, however, $\theta_{SH}$ of Ti is negligible (–0.00036) (ref. [26]) while Pt is known as a material which has high $\theta_{SH}$ (~0.1) (ref. [27,28]), so even considering the high resistivity of Ti, the larger SCC signal observed cannot be properly explained. As a further check, we also fabricated devices with Ta as the NM. The sign of $\theta_{SH}$ in Ta is opposite to the one of Pt, but the SCC signal of a BiSe(2 nm)/Ta(10 nm) device has the same sign as the other devices (Note S2 in the Supporting Information), clearly showing that the SCC appears in the thin BiSe layer, regardless of the NM used.

Figure 2e shows $2\Delta R_{SCC}$ as a function of the BiSe thickness in the BiSe (2 to 16 nm)/Ti(10 nm) local spin injection devices. $2\Delta R_{SCC}$ is the largest for 3 nm of BiSe, and it decreases for thicker structures. The same experiment was conducted with BiSe (3 to 5 nm)/Pt(10 nm), as shown in Figure 2f, yielding a similar trend. The SCC signal could only be observed up to 16 nm of BiSe for Ti and up to 5 nm for Pt, with no signal obtained at a thickness beyond that, mostly due to the increasingly larger noise level and low SCC signal.

We should mention that an anomalous Hall effect (AHE)[7,29,30] and ordinary Hall effect (OHE)[31] can appear as artifacts in the measured SCC signals (Note S3 in the Supporting Information). The anomalous Hall angle obtained by applying an out-of-plane magnetic field to CoFe in our own devices is 1.5 %. In the local spin injection device, the contribution of the AHE is greatly reduced since the magnetization points along the $x$-axis and only the contribution by the current flowing in the $z$-axis is considered. The contribution of the AHE was calculated by 3D finite element method (FEM) based on the spin diffusion model[7,29,30,32], and we obtained less than 6 mΩ and 2 mΩ AHE signals in the whole thickness range for BiSe/Ti and BiSe/Pt structures, respectively. Additionally, the stray field of CoFe-induced OHE is calculated by 3D FEM simulation based on the Hall coefficient of BiSe/NM structures obtained by applying the out-of-plane magnetic field to each structure[31]. We obtained less than 0.3 mΩ OHE contribution for both structures. As shown in Figure 2e and 2f, the calculated AHE and OHE signals decay with BiSe thickness since current shunting is suppressed and the distance between FM and NM increases, respectively.

Next, we analyzed the thickness dependence of the SCC signal by 3D FEM simulation[7] also taking into account the AHE and the OHE discussed above (see Note S4 in the Supporting Information for details). The simulation is performed by assuming that the resistivity of BiSe is 18,000 μΩ·cm (20-nm-thick and 80-nm-wide wire in Figure 1d). The resistivity of CoFe and Ti were 42 and 40 μΩ·cm, respectively, which were measured directly on the device. The obtained spin diffusion length ($\lambda_s$) is 0.5 nm and $\theta_{SH}$ is 27.5 (see Figure S4d), a very large value which is in a good agreement with the previously reported value, 18.62 (ref. [21]), and it seems to prove once again that sputtered BiSe is one of the most promising material for SCC devices, in particular the MESO logic device, which also requires a high resistivity.

However, unlike the previous report, the quantum confinement[21] cannot be confirmed here, as a single $\theta_{SH}$ value of 27.5 is obtained over the whole BiSe thickness range of 3−16 nm. The hybridization of topological surface state (TSS) reduces the SCC signal when a TI is thinner than 6 nm[33], but the maximum signal is at 3 nm as shown in Figure 2e. The activation of TSS or the suppression of bulk conduction at cryogenic temperature reduces the resistivity[34,35], which is not observed in this work as shown in Figure 1f. From all these facts, it is reasonable to consider sputtered polycrystalline BiSe as a normal conductor rather than a TI.

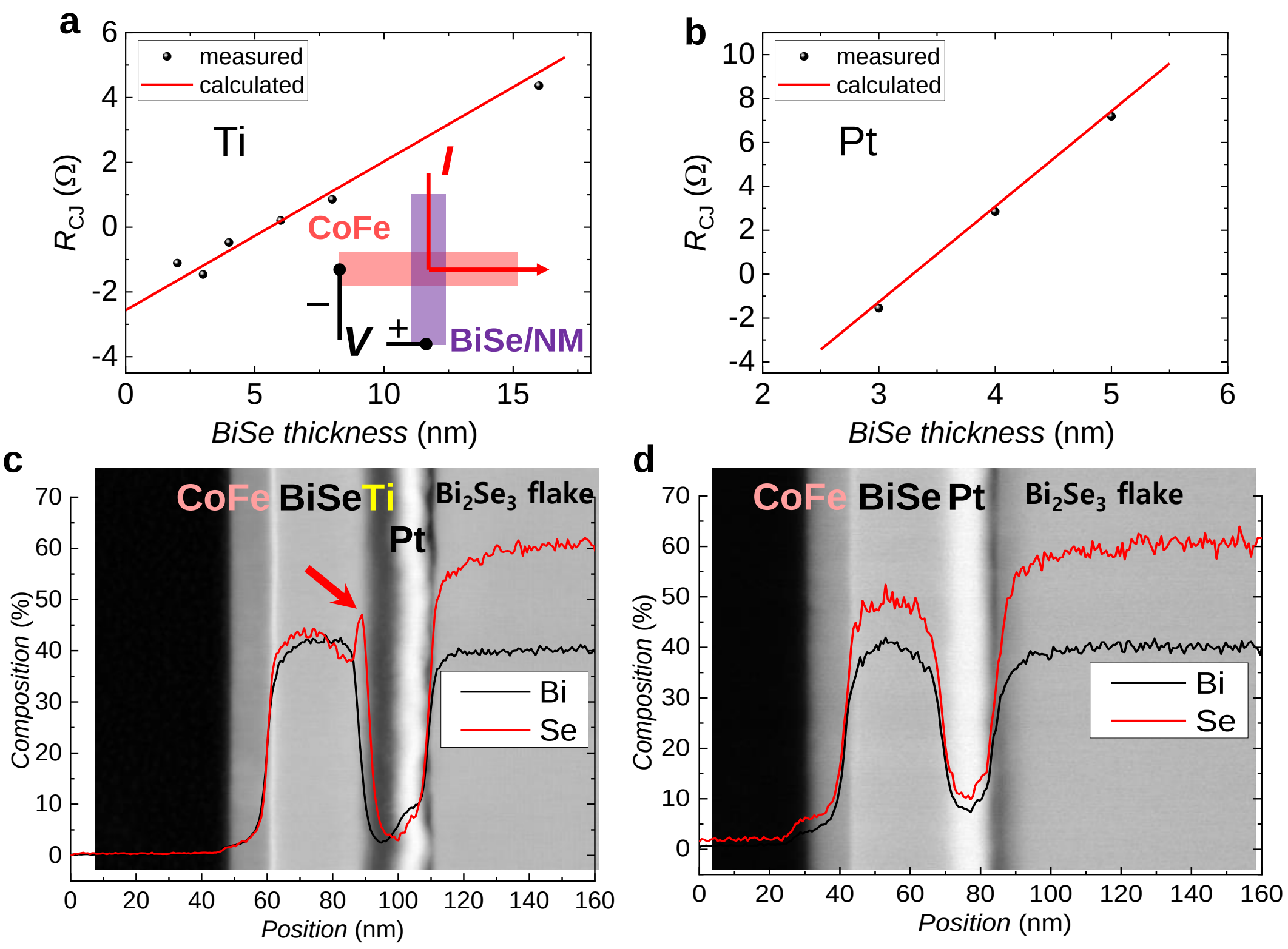

**Figure 3.** Estimation of BiSe resistivity in vertical direction. Cross-junction resistance of (a) BiSe/Ti and (b) BiSe/Pt structures. Inset: measurement geometry. TEM images and EDX scan of (c) CoFe/BiSe/Ti and (d) CoFe/BiSe/Pt thin film structures compared with single crystal $Bi_2Se_3$ flakes exfoliated on top for a proper quantification of the composition of the sputtered BiSe layers.

On the other hand, the cross-junction resistance ($R_{CJ}$) as a function the thickness of BiSe in Figure 3a and 3b disagree with such a high resistivity of BiSe. The sketch in Figure 3a shows the measurement configuration of $R_{CJ}$. In the BiSe/Ti structures shown in Figure 3a, $R_{CJ}$ shows a constant increase with the BiSe thickness from 2 to 16 nm, but the overall values are unexpectedly low. Even in BiSe/Pt devices shown in Figure 3b, although the change of $R_{CJ}$ is higher than that of BiSe/Ti, the values are rather similar. This measurement assumes a vertical current flowing uniformly across the junction. Considering the large resistivity difference between BiSe and metallic wires such as Pt, Ti, and CoFe, the current stays in the CoFe until reaching the junction and flows vertically through BiSe to NM, so that the $R_{CJ}$ is expected to be more than 80 Ω for 3-nm-thick BiSe when a resistivity of 18,000 μΩ·cm and a junction area of 80 nm × 80 nm are considered. However, the measured $R_{CJ}$ on the devices with BiSe thickness up to 4 nm has negative values, which is generally considered as a transparent interface, hinting that the resistivity of BiSe may be lower than initially measured in nanostructures without the NM. In order to reliably extract the resistivity of BiSe, we perform a 3D FEM simulation[36] of the cross-junction resistance measurements, where the independently measured resistivities of the NM (Ti or Pt) and the CoFe are used as input and the only unknown parameter is the BiSe resistivity. The 3D FEM simulation results are shown by red lines in Figure 3a and 3b. The resistivity of BiSe is calculated to be 600 μΩ·cm for the BiSe/Ti structure, which is 30 times smaller than the values measured in the BiSe wire in Figure 1d, and 3,700 μΩ·cm in the BiSe/Pt structure, about 6 times higher than that of BiSe/Ti. These results indicate that the top NM layer changes the resistivity of BiSe. In the following paragraphs, we will discuss the origin of such variation and the consequences in the quantification of the spin Hall effect in sputtered BiSe.

To understand the relationship between BiSe resistivity and the NM used, we performed TEM and EDX experiments in $SiO_2$ substrate/CoFe(15 nm)/BiSe(30 nm)/Ti(10 nm)/Pt(10 nm) and $SiO_2$ substrate/CoFe(15 nm)/ BiSe(30 nm)/Pt(10 nm) multilayer stacks as shown in Figure 3c and 3d (see also Note S5 in the Supporting Information for the analysis of the actual BiSe/Ti local spin injection device). In order to obtain a quantitative elemental analysis from the EDX, we exfoliated single-crystal $Bi_2Se_3$ flakes on top of each stack. Since the $Bi_2Se_3$ flakes are stoichiometric, the composition of sputtered BiSe can be accurately determined by normalizing the Bi and Se intensity curves to 40% and 60%, respectively. In the BiSe/Ti structure, a clear peak of the Se curve, indicated by the red arrow in Figure 3c, appears as a result of strong intermixing near Ti[22–25], and BiSe has a 50:50 composition in the rest of the layer. No intermixing was observed between the BiSe and Pt, where the composition was 45:55. Eventually, we obtained two different BiSe compositions depending on the chosen NM, even with the same growth condition. This is the reason why the resistivity of BiSe varied from 600 to 3,700 μΩ·cm in the two structures depending on the NM used. The critical role of intermixing in SCC is further confirmed with harmonic Hall measurements (Note S6 in the Supporting Information).

Still, the composition change alone cannot explain the high resistivity (18,000 μΩ·cm) obtained in the BiSe wire measured laterally (Figure 1f). Using impedance measurements, we confirmed that there are two resistance elements, grain and grain boundary (Note S7 in the Supporting Information). Accordingly, the absence of grain boundaries in the vertical direction for BiSe thickness below 16 nm drastically reduces the BiSe resistivity when measured along this direction as compared to the lateral measurement, which includes the grain boundary contribution making high noise level. The BiSe wire has a resistivity of 18,000 μΩ·cm as a series resistance of grain and grain boundary without intermixing, and the resistivity of 3,700 μΩ·cm measured vertically in the BiSe/Pt structure is of the parallel resistance of grain and grain boundary without intermixing. On the other hand, in the BiSe/Ti structure, the resistivity of 600 μΩ·cm appears as a parallel resistance of grain and grain boundary accompanied by composition change due to intermixing. The different conditions are summarized in Table 1.

**Table 1. Summary of BiSe resistivities depending on measurement direction and NM**

| Structure | Measurement direction | Resistance of grain & boundary | Intermixing | Resistivity [μΩ·cm] |
|---|---|---|---|---|
| BiSe wire | Lateral | In series | No | 18,000 |
| BiSe/Pt | Vertical | In parallel | No (Se 55 %) | 3,700 |
| BiSe/Ti | Vertical | In parallel | Yes (Se 50 %) | 600 |

Returning to the SCC results of Figure 2e and 2f, it is necessary to accurately estimate $\theta_{SH}$ once again, because the resistivity of BiSe can affect the spin injection efficiency and current shunting, and consequently $\theta_{SH}$. As a result of the 3D FEM simulation performed using the resistivity obtained in Figure 3a and 3b, $\lambda_s$ of 0.5 nm and $\theta_{SH}$ of 0.45 are estimated for the BiSe/Ti devices. In the case of BiSe/Pt devices, we extract a $\lambda_s$ of 0.35 nm and $\theta_{SH}$ of 3.2. These results are significantly different from previously reported values[21,37] and our first estimation, because the high BiSe resistivity reduces the SCC signal by strong spin back flow. In addition, the SCC signal reduction as BiSe gets thicker is completely explained by spin diffusion model for the entire thickness ranges, indicating the absence of quantum confinement depending on the grain size[21].

**Table 2. Summary of spin-related parameters in this work and previous reports**

| | BiSe thickness [nm] | $\rho_{BiSe}$ [μΩ·cm] | $\lambda_s$ [nm] | $\theta_{SH}$ | $\lambda_{IREE}$ [nm] | Method | FM |
|---|---|---|---|---|---|---|---|
| $Bi_{50}Se_{50}$/Ti (this work) | 2 - 16 | 600 | 0.5 | 0.45 | 0.225* | Electrical spin injection | CoFe |
| $Bi_{45}Se_{55}$/Pt (this work) | 3 - 5 | 3,700 | 0.35 | 3.2 | 1.12* | | |
| Ref. 21 | 4 - 40 | 12,820 | - | 18.62† | - | harmonic Hall (d.c.) | CoFeB |
| Ref. 40 | 2 - 16 | - | - | - | 0.32† | Spin pumping | CoFeB |
| Ref. 41 | 4 - 16 | - | - | - | 0.11† | Spin pumping | YIG |
| Ref. 37 | 5 - 10 | 1,000,000 | - | 75 | - | ST-FMR‡ | Py |
| Ref. 38 | 3 - 15 | 890 | - | 0.35 | - | harmonic Hall (a.c.) | Co |

* $\lambda_{IREE} = \theta_{SH} \cdot \lambda_s$.
†The maximum value of the thickness dependence is indicated.
‡Spin torque-ferromagnetic resonance.

By comparing our results with the reported results shown in Table 2, it is possible to strictly judge the SCC efficiency of sputtered BiSe. The BiSe thickness range used in this research covers all references in Table 2. The resistivity obtained in papers differs up to 1000 times[37,38] because of different growth conditions and measurement techniques. Only Ref. 38 reports a resistivity similar to that of the BiSe/Ti structure with also a similar $\theta_{SH}$, in agreement with our claim. To compare with the results of spin pumping experiments, in which the $\lambda_{IREE}$ is used to quantify the SCC efficiency, the product $\lambda_s \cdot \theta_{SH}$ which is more relevant to SCC efficiency in local spin injection device is considered[7,28,39]. In particular, the $\theta_{SH}$ of 18.62 (ref.

[21]) is more than 50 times higher than 0.35 which is obtained in the BiSe/Ti structure, but $\lambda_{IREE}$ of 0.32 nm (ref. [40]) and 0.1 nm (ref. [41]) are comparable to what we obtain in our local spin injection devices using BiSe/Ti (0.225 nm) despite differences in experimental methods and ferromagnetic materials. These results indicate that intermixing must be considered in material characterization, especially an appropriate resistivity quantification, and that $\theta_{SH}$ can be overestimated when these aspects are not properly considered.

To conclude, we observed all electrical spin-to-charge conversion in sputtered $Bi_xSe_{1-x}$ in local spin injection devices at room temperature and showed that all parameters related to SCC efficiency which are resistivity, $\lambda_s$ and $\theta_{SH}$ were affected by intermixing with the adjacent non-magnetic metal used to electrically shunt $Bi_xSe_{1-x}$. In particular, the fact that Se concentration change by intermixing made a difference of 6 times in resistivity shows how easily $\theta_{SH}$ can be overestimated by resistivity without considering intermixing. Even though the SCC signal obtained in this study is too small to realize a MESO logic device, it allowed us to quantify the SCC efficiency of sputtered $Bi_xSe_{1-x}$ in functional spintronic devices (instead of commonly used $Bi_xSe_{1-x}$/FM bilayers). The potential of highly resistive sputtered $Bi_xSe_{1-x}$ as the active element in the reading module can be exploited by improving different aspects, such as reducing the electrical noise caused by the grain boundary so that the top NM layer shunting the signal can be removed. The NM layer was applied to reduce noise level in electrical measurement, but it is not a fundamental solution due to its low resistivity, leading to an overall SCC signal reduction. We suggest that the SCC efficiency and the SCC signal magnitude can be increased by protecting $Bi_xSe_{1-x}$ with a tunneling barrier, such as MgO, to prevent intermixing, current shunting, and spin back flow, while enhancing the spin injection efficiency.

## METHODS

**Device nanofabrication.** The devices were fabricated on Si/$SiO_2$ substrates by e-beam lithography, sputter deposition, ion-milling and lift-off process. The CoFe wire is patterned as the first layer and deposited by dc magnetron sputtering of 30 W at 2 mTorr of Ar pressure. To eliminate a sidewall of CoFe, ion-milling (Ar flow of 15 sccm, acceleration valtage of 50 V, beam current of 50 mA, and beam voltage of 300 V) is performed with incident beam angle of 80° after lift-off process. The T-shape wire is patterned, the BiSe layer is deposited by magnetron sputtering of $Bi_2Se_3$ target using rf power of 35 W at 3 mTorr, and subsequently NM is in-situ deposited using dc power (80 W for Pt, 100 W for Ti, 200 W for Ta) at 3 mTorr.

**Materials characterization.** TEM, STEM and EDX data were obtained on (Scanning) Transmission Electron Microscope Titan G2 60-300 (FEI, Netherlands) at 300kV accelerating voltage. The Microscope was equipped with high brightness xFEG, imaging Cs corrector, STEM HAADF detector and EDX RTEM spectrometer (EDAX, UK). Cross-sectional TEM samples of the devices and multilayers were prepared by a standard FIB technique.

**Electrical measurements.** Electronic transport measurements are performed in a Physical Property Measurement System (PPMS) from Quantum Design, using a 'd.c. reversal' technique with a Keithley 2182 nanovoltmeter and a 6221 current source. The second harmonic measurement is performed in PPMS using AC Transport Measurement System. Magnetic field is applied by a superconducting solenoid magnet, and a rotatable sample stage is used.

## ASSOCIATED CONTENT

**Supporting Information**.

The Supporting Information is available at the end of this document.

Further information about TEM and EDX characterization of additional stacks, spin-to-charge conversion in Ta local spin injection device, quantification of spurious effects (AHE of CoFe and OHE of Pt), spin diffusion 3D FEM simulation for spin-to-charge conversion, impedance measurements, and second harmonic measurement of spin torque.

## AUTHOR INFORMATION

### Corresponding Author

*E-mail: w.choi@nanogune.eu, f.casanova@nanogune.eu

### Author contributions

†These authors contributed equally: Won Young Choi, Isabel C. Arango.

### Notes

The authors declare no competing interests.

## ACKNOWLEDGEMENTS

We acknowledge C. Rufo, R. Llopis and R. Gay for technical assistance with the sample fabrication. This work is supported by Intel Corporation through the Semiconductor Research Corporation under MSR-INTEL TASK 2017-IN-2744 and the 'FEINMAN' Intel Science Technology Center, and by the Spanish MICINN under project No. RTI2018-094861-B-I00 and under the Maria de Maeztu Units of Excellence Programme (MDM-2016-0618 and CEX2020-001038-M). W.Y.C. acknowledges postdoctoral fellowship support from 'Juan de la Cierva Formación' programme by the Spanish MICINN (grant No. FJC2018-038580-I). D.C.V. acknowledges support from the European Commission for a Marie Sklodowska-Curie individual fellowship (Grant No. 892983-SPECTER).

# *SUPPORTING INFORMATION*

**Note S1**

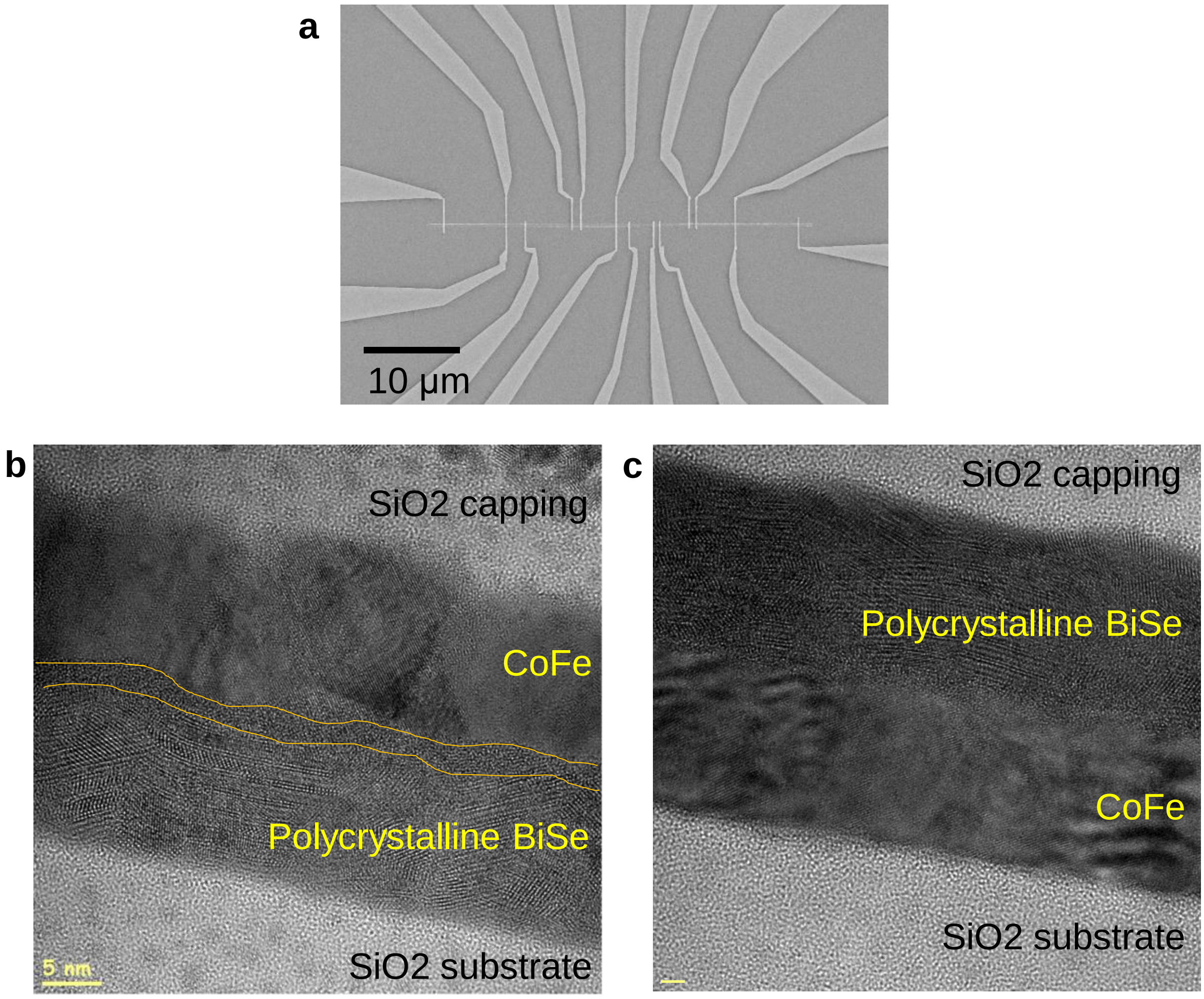

**Figure S1.** (a) SEM image of a BiSe wire (horizontal) and multiple metallic contacts (vertical) with various distances used for Figure 1 (e,f) in the main text. (b,c) High-resolution TEM images of (b) BiSe/CoFe, and (c) CoFe/BiSe structures.

The BiSe wire has not only a high resistivity but also a high noise level in electrical measurements. Due to a big difference in resistivity between grain and boundary (see Figure S6), internal capacitance affects electrical measurements making them unstable. To stabilize the measurement and reduce the noise level, a NM (Ti, Pt or Ta) is deposited on top of BiSe for local spin injection devices. It largely reduces the current path inside BiSe, which means that the electrical measurement is less affected by the noise source, efficiently reducing the noise level. Why BiSe itself is a noise source is discussed in Supplementary Section 6.

Figure S1b and S1c show polycrystalline BiSe and intermixed layer depending on its stacking order. While the BiSe/CoFe structure in Figure S1b has a thick amorphous intermixed layer, the CoFe/BiSe in Figure S1c has a vague one. In some grains, alternating $Bi_2Se_3$ quintuple layers and Bi bilayers are found, but its frequency varies even within a grain.

**Note S2**

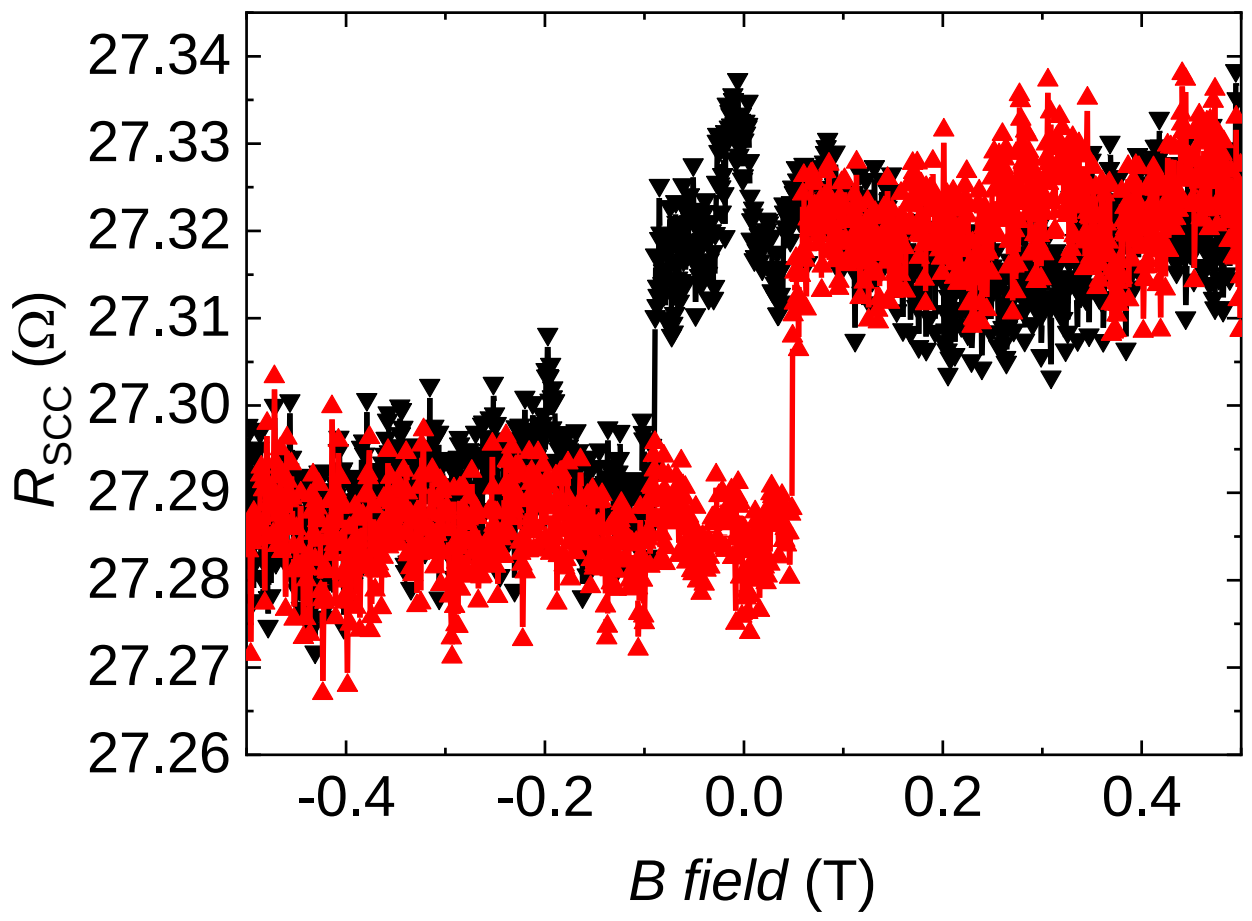

**Figure S2.** SCC resistance as a function of the magnetic field obtained in a BiSe/Ta local spin injection device.

To confirm whether the SCC signal originates from the BiSe or the NM, we prepared a local spin injection device with BiSe(2 nm)/Ta(10 nm) in the T-shaped nanostructure. Since the SCC signal of Ta is expected to be opposite to BiSe and Pt, if SCC in the NM is dominant, the polarity of the SCC signal in BiSe/Ta device will be opposite. The obtained signal magnitude is about 30 mΩ and it is larger than other Ti and Pt devices with similar BiSe thickness, because of a high resistivity of Ta, 180 μΩ·cm. Although BiSe is only 2 nm thick in the device, the observed SCC signal has the same polarity as the other devices (BiSe/Pt and BiSe/Ti), confirming that SCC in BiSe is dominant in our devices.

**Note S3**

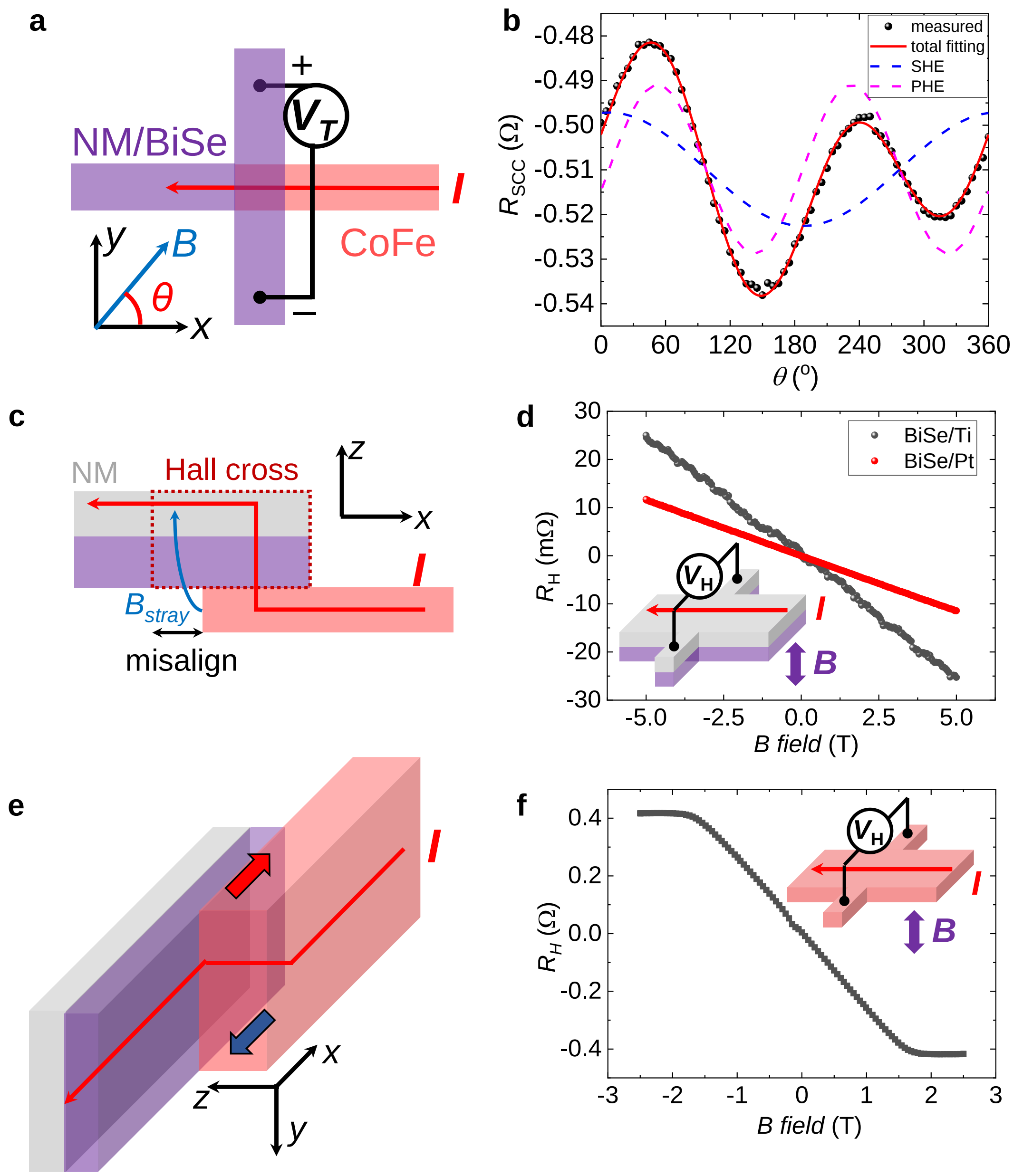

**Figure S3.** Different contributions present in a local spin injection device. Schematics and signals of (a,b) planar Hall effect, (c,d) ordinary Hall effect, and (e,f) anomalous Hall effect.

Artifacts that can be observed with the SCC measurement include planar Hall effect (PHE), ordinary Hall effect (OHE), and anomalous Hall effect (AHE). First, to measure PHE using a local spin injection device, an in-plane magnetic field (3 T) was applied, and an angle scan was performed as shown in Figure S3a. The result is shown in Figure S3b. This measurement is composed of the sum of the two elements, SCC (due to the SHE) and PHE, which have an angular dependence following $cos\,\theta$ and $sin\,2\theta$, respectively. Since the SCC signal is measured by applying an x-axis magnetic field, the amplitude of the cosine term becomes $2\Delta R_{SCC}$. On the other hand, in the case of PHE, since the period is $2\theta$, there is no contribution as a signal in the same measurement. Although a baseline change due to misalignment between the injected current and the magnetic field (phase shift) can appear, it does not modify the value of $2\Delta R_{SCC}$ extracted from saturation at positive and negative magnetic fields, as discussed extensively in Ref. *30*.

Second, OHE is mainly induced by the stray field ($B_{\text{stray}}$) in z-axis generated at the tip of the CoFe wire, as shown in Figure S3c. Since $B_{\text{stray}}$ acts strongly up to about 20 nm from the CoFe tip, the simulation was performed considering the misalignment of 20 nm in the 3D FEM model. In an actual device, the misalignment is less than 20 nm, but this was considered in order not to underestimate the contribution. $B_{\text{stray}}$ can have both x- and z-axis components, and since the current at the junction has a z-axis component, the x-axis component of $B_{\text{stray}}$ must also be taken into account. The external magnetic field is also applied in the x-axis, so it is expected to be observed with a slope in the SCC signal, but the observed signal is almost flat. Therefore, $B_{\text{stray}}$ in x-axis is not considered. To obtain the Hall coefficient, a Hall bar of BiSe (16 nm)/NM (Ti or Pt 10 nm) was prepared, and the Hall effect was observed by applying an out of plane magnetic field as shown in Figure S3d. Due to the large resistivity difference between BiSe and NM, most of the current flows through the NM, and the measurement result is totally dependent on NM. The same applies to the local spin injection device, and since the distance between NM and CoFe changes according to the thickness of BiSe, OHE also has a thickness dependence. The calculated OHE contribution in the entire thickness range of BiSe does not exceed 0.5 mΩ, and the results are shown together with AHE in Figure 2e and 2f of the main manuscript.

Finally, AHE acts as an artifact as the potential difference generated in the CoFe wire is transmitted to the NM. Considering the x-axis magnetization, AHE can occur only by the current in z-axis at the junction as shown in Figure S3e. Figure S3f is the AHE signal of the CoFe wire, and the anomalous Hall angle ($\theta_{AH}$) is 1.5% when CoFe is 15-nm-thick, and its resistivity is 42 $\mu\Omega \cdot cm$. The AHE contribution decreases with the thickness of BiSe because BiSe acts as a barrier. The 3D FEM simulation result is shown in Figure 2e and 2f of the main text.

In addition, it is possible to reduce the OHE and AHE contributions by changing the design of the device. Since the stray field is proportional to the area of the tip of the ferromagnet , making CoFe wire as thin and narrow as possible will minimize the OHE contribution. Extending the tip of the FM beyond the junction area, while isolating it from the middle arm of the T-structure, is another possibility, but more challenging from the nanofabrication perspective. In the case of AHE, a previous work by our group[30] showed that the relative thickness of the FM and the spin-orbit material can be tuned to completely remove the AHE contribution. The optimal thicknesses will also depend on the resistivity of each material. Inserting a MgO barrier between CoFe and BiSe could also be an effective way to reduce both OHE and AHE contributions. In the OHE case, MgO barrier plays a role of spacer, so the stray field at BiSe/NM will decrease. In the AHE case, since the anomalous Hall voltage induced in CoFe will be blocked by the barrier, it will not be transferred to the BiSe/NM wire.

**Note S4**

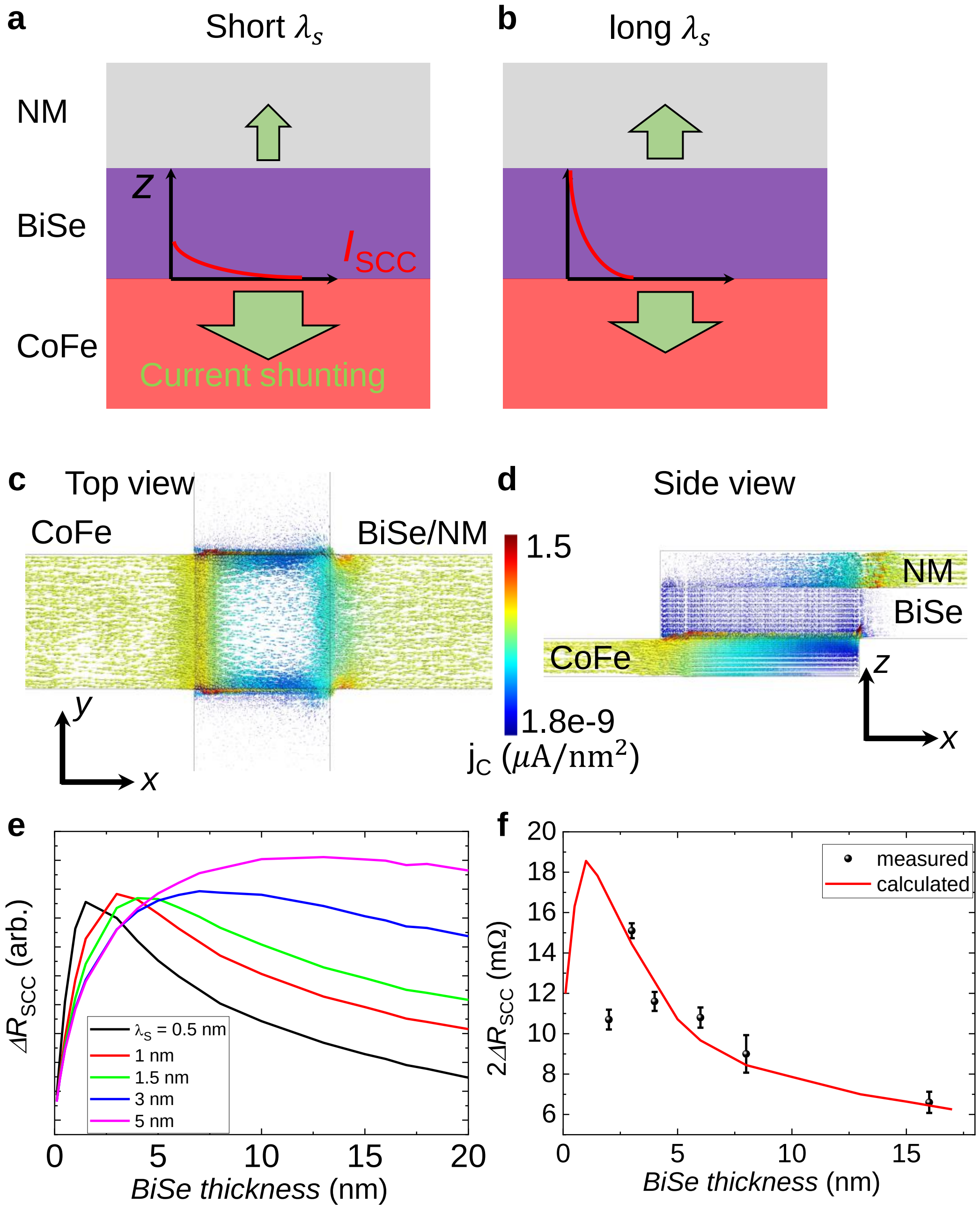

**Figure S4.** Illustration of the shunting of the converted charge current ($I_{SCC}$) into the CoFe and NM layers when $\lambda_s$ is (a) shorter than, and (b) comparable to the thickness of BiSe layer. Local current density obtained from the 3D FEM simulation in (c) top view, and (d) side view. 3D FEM simulation results of (e) $\lambda_s$ dependence, and (f) BiSe thickness dependence of SCC signal observed in BiSe/Ti local spin injection devices.

FEM simulations[7,29,30,32] are performed within the framework of the two-current drift-diffusion model, with the collinear magnetization of the FM electrode along the easy axis. The geometry construction and 3D-mesh were elaborated using the free software GMSH with the associated solver GETDP[42] for calculations, post-processing and data flow control.

The fitting parameters of 3D FEM simulation in this model are the spin Hall angle ($\theta_{SH}$) and the spin diffusion length ($\lambda_s$) of BiSe while spin polarization, anomalous Hall angle and magnetization of CoFe, Hall coefficient of NM and resistivities of wires are fixed parameters. The spin current injected into BiSe will be converted into a transverse charge current ($I_{\mathrm{SCC}}$) proportional to $\theta_{SH}$ and the amount of local $I_{\mathrm{SCC}}$ will decay with $\lambda_s$, sketched as a red line in Figure S4a and S4b. Detailed local current density is shown in Figure S4c and S4d. When $\lambda_s$ is much shorter than the BiSe thickness, the $I_{\mathrm{SCC}}$ is concentrated near the interface between CoFe and BiSe, so it easily flows back into CoFe, reducing the SCC signal which is measured along the BiSe/NM wire. On the other hand, when $\lambda_s$ becomes comparable to the BiSe thickness, the distribution of $I_{\mathrm{SCC}}$ over BiSe thickness will span over the entire thickness, increasing current shunting to the NM layer. Therefore, the measured SCC signal depends on BiSe thickness and $\lambda_s$ as shown in Figure S4e, whereas $\theta_{SH}$is related to the overall signal magnitude. When $\lambda_s$ and BiSe thickness are similar, the signal magnitude is proportional to the resistivity of NM because what we measure is the voltage. Finally, we can simultaneously obtain $\lambda_s$ and $\theta_{SH}$ by performing the 3D FEM simulation.

Figure S4f shows the result of 3D FEM simulation for SCC signal in BiSe/Ti local spin injection device using a BiSe resistivity of 18,000 μΩ·cm. The parameters used for the model are resistivity of CoFe (42 μΩ·cm), Ti (40 μΩ·cm), spin polarization of CoFe (0.48), and spin diffusion length of CoFe, which is calculated by $\rho_{\mathrm{CoFe}} \cdot \lambda_{\mathrm{CoFe}}$=1.29 fΩ·m$^2$ (ref. [43]). Because of such a high resistivity of BiSe, the charge current shunting to the NM layer is highly suppressed in the simulation, thus $\theta_{SH}$ becomes high, 27.5, to compensate for the suppression of this shunting. $\lambda_s$ is estimated to be 0.5 nm, which is 4 times shorter than the thinnest BiSe layer used in our experiment, so we consider that SCC occurs only in BiSe.

On the other hand, the 3D FEM simulation has a limitation. The simulation of the main text shows the results considering both AHE and OHE, but it was modeled under one important assumption, which is the absence of the intermixed layer. Figures 1a–d, 3c and 3d show the intermixing occurring at the interfaces of CoFe/BiSe and BiSe/NM. The intermixed layers are at the CoFe/BiSe and BiSe/Ti interfaces and are generated by diffusion of Se atoms. As a result, layers composed of CoFe-Se and Ti-Se are formed, and the Se concentration of BiSe decreases by the amount of Se escaped. The resistivity of BiSe decreases due to the change in the Se concentration, and the resistivity of the intermixed layer is expected to increase. Since Se diffusion is not observed, this is not the case at the BiSe/Pt interface.

The local spin injection device has a CoFe/BiSe/NM structure (two interfaces), but since there is no practical way to obtain all parameters for 3D FEM model such as resistivity, spin diffusion length, and spin polarization of the interface, these interfaces cannot be considered. On the other hand, in general, the higher the interface resistance, the more evenly the current flows through layers, and the shunting of the converted charge current is suppressed. Also, the CoFe/BiSe interface can affect the spin injection efficiency. Formation of the CoFe-Se layer can lower the spin polarization of the FM interface and increase the spin injection efficiency by reducing the spin back flow by increasing the interface resistance. However, considering the high BiSe resistivity, such effects are not expected to be significant.

The presence of intermixed layers can also affect the resistivity evaluation of each layer. For example, the resistivity of Ti was obtained from a BiSe/Ti bi-layered wire, and since BiSe resistivity is much higher than that of Ti, most of the current flows through Ti and the resistivity of the bi-layered wire is almost the same as that of Ti. However, this assumption is incomplete because it does not consider intermixed layers which can effectively increase the Ti resistivity as the intermixed layer thickens. Since information of their resistivity cannot be obtained experimentally, intermixed layers are not considered in the 3D FEM model.

**Note S5**

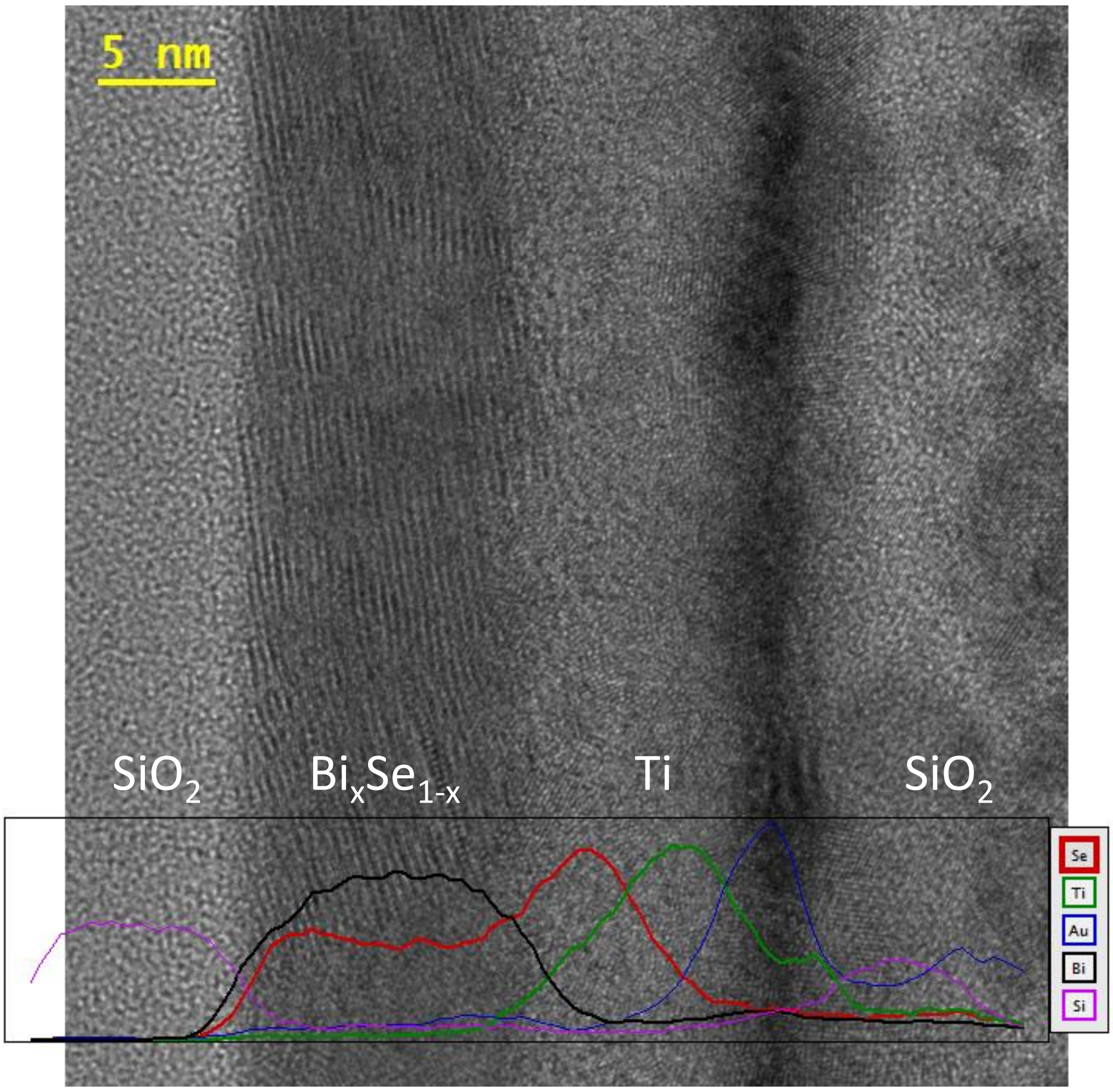

**Figure S5.** TEM image and EDX curves of BiSe/Ti structure.

Figure S5 shows the TEM image and EDX curves similar to Figure 3c in main text, but with 16-nm-thick BiSe. These are obtained by cutting the BiSe/Ti T-shape wire of the local spin injection device. To know the exact composition of BiSe, we need a BiSe flake to normalize as shown in Figure 3c, but even without the flake, Se diffusion can be clearly seen.

## Note S6

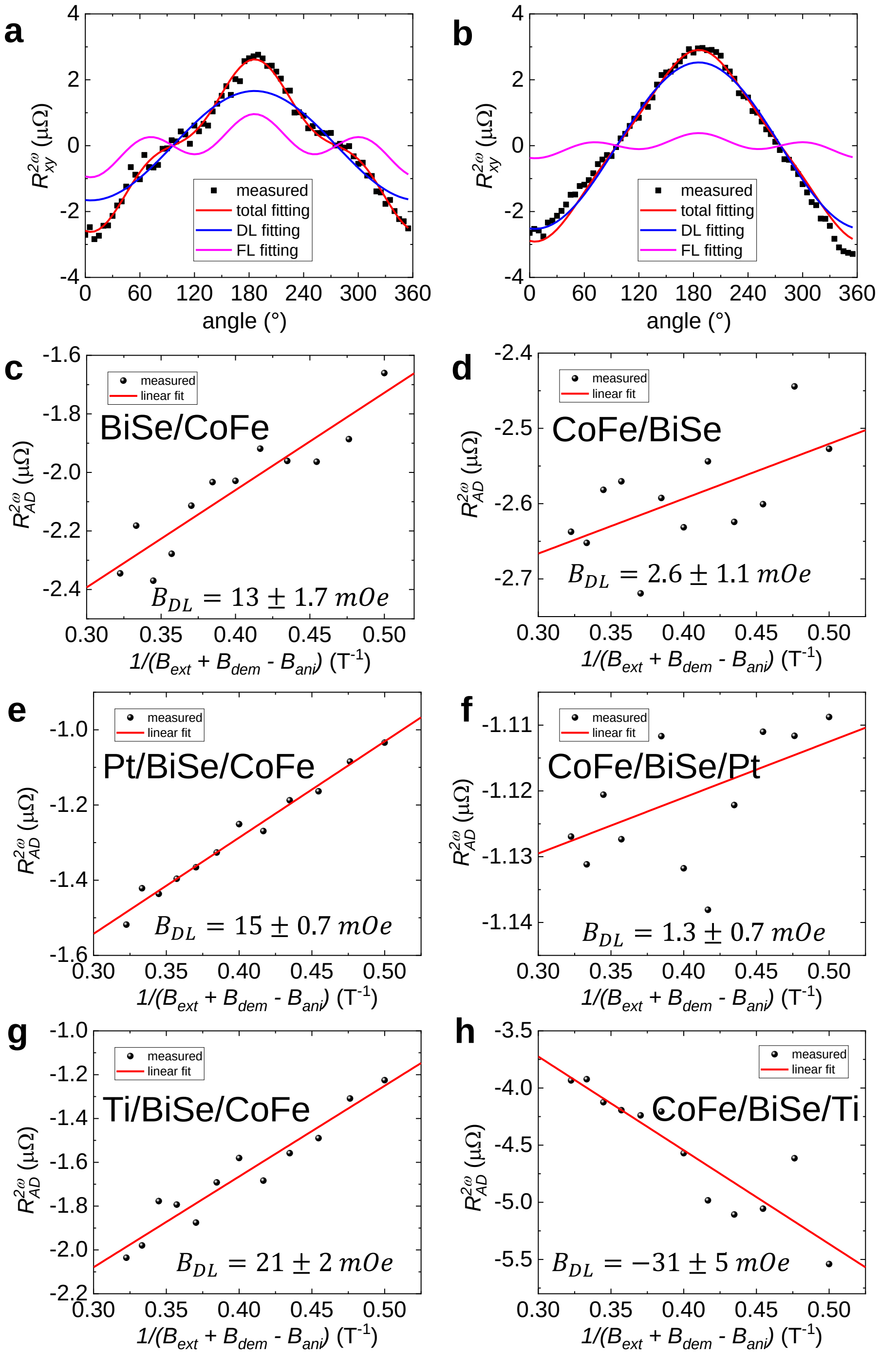

**Figure S6.** Second harmonic measurement of spin torque. Raw data and fitting curves using Eq. 1 of (a) BiSe/CoFe, and (b) CoFe/BiSe structures under 0.1 T. The second harmonic resistance induced by damping-like torque of (c) BiSe/CoFe, (d) CoFe/BiSe, (e) Pt/BiSe/CoFe, (f) CoFe/BiSe/Pt, (g) Ti/BiSe/CoFe, and (h) CoFe/BiSe/Ti stacks with the damping-like torque field, $B_{DL}$, obtained by linear fitting.

For further study on intermixing, 5 $\mu m$ wide Hall bar devices with various structures are prepared. 3-nm-thick CoFe, 4-nm-thick BiSe and 2-nm-thick NM are used, then capped with $SiO_2$. Figure S5 shows the results of second harmonic measurement of spin torque[44] for each structure. For the measurement, an alternating current of 3 mA is applied, and in-plane magnetic field is applied from 0.1 T to 1.2 T. Figure S6a and S6b are the results of second harmonic measurement under 0.1 T of BiSe/CoFe and CoFe/BiSe, respectively. The fitting curves are obtained using the equation[44],

$$R_{xy}^{2\omega} = \left(R_{AHE}\frac{B_{DL}}{B_{ext}} + I_0\alpha\nabla T\right)\cos\varphi + 2R_{PHE}(2\cos^3\varphi - \cos\varphi)\frac{B_{FL}+B_{Oe}}{B_{ext}}, \qquad (1)$$

where φ is azimuthal angle, $R_{\mathrm{AHE(PHE)}}$ is signal magnitude of AHE(PHE), $B_{\mathrm{DL(FL)}}$ is damping(field)-like torque field, $B_{\mathrm{Oe}}$ is Oersted field, $I_0$ is amplitude of a.c. current, $\nabla T$ is thermal gradient, $\alpha$ is coefficient of anomalous Nernst effect (ANE). $R_{\mathrm{AHE}}$ is measured by out-of-plane magnetic field scan, and $R_{\mathrm{PHE}}$ is determined by the first harmonic angle scan under in-plane magnetic field. Then, the results are decomposed into two terms, damping-like and field-like torque. Figure S6c and S6d show the second harmonic transverse resistances induced by damping-like torque and ANE, and its linear fit give us $B_{\mathrm{DL}}$ as an indicator of SCC. The intermixing between CoFe and BiSe is strong in the top CoFe layer (left panel) and weak in the bottom CoFe layer (right panel), as we discussed in the main text (Figure 1c and 1d). Accordingly, the BiSe layer of BiSe/CoFe structure has relatively low resistivity, and more current flows through the layer. Since the stacking order of the two structures is opposite, the slope of the fitting line and damping-like torque field ($B_{\mathrm{DL}}$) must also have opposite sign. However, in the case of the CoFe/BiSe structure, the fitting line has a positive slope, and the data points are scattered, so it is not the expected result in ideal conditions. This issue is also observed in Figure S6e and S6f observed in structures including Pt. Note that the sign of $\theta_{SH}$ does not change due to stacking order, which is confirmed by measurement of local spin injection device with opposite stacking order. On the other hand, Ti strongly intermixes with BiSe and dramatically lowers the resistivity of BiSe, so both structures with opposite stacking order show reasonable results such as a slope sign of fitting line and a signal to noise ratio (Figure S6g and S6h). As a result, the difference in resistivity due to intermixing becomes an important factor in determining whether the signal of the second harmonic measurement is measurable or not.

## Note S7

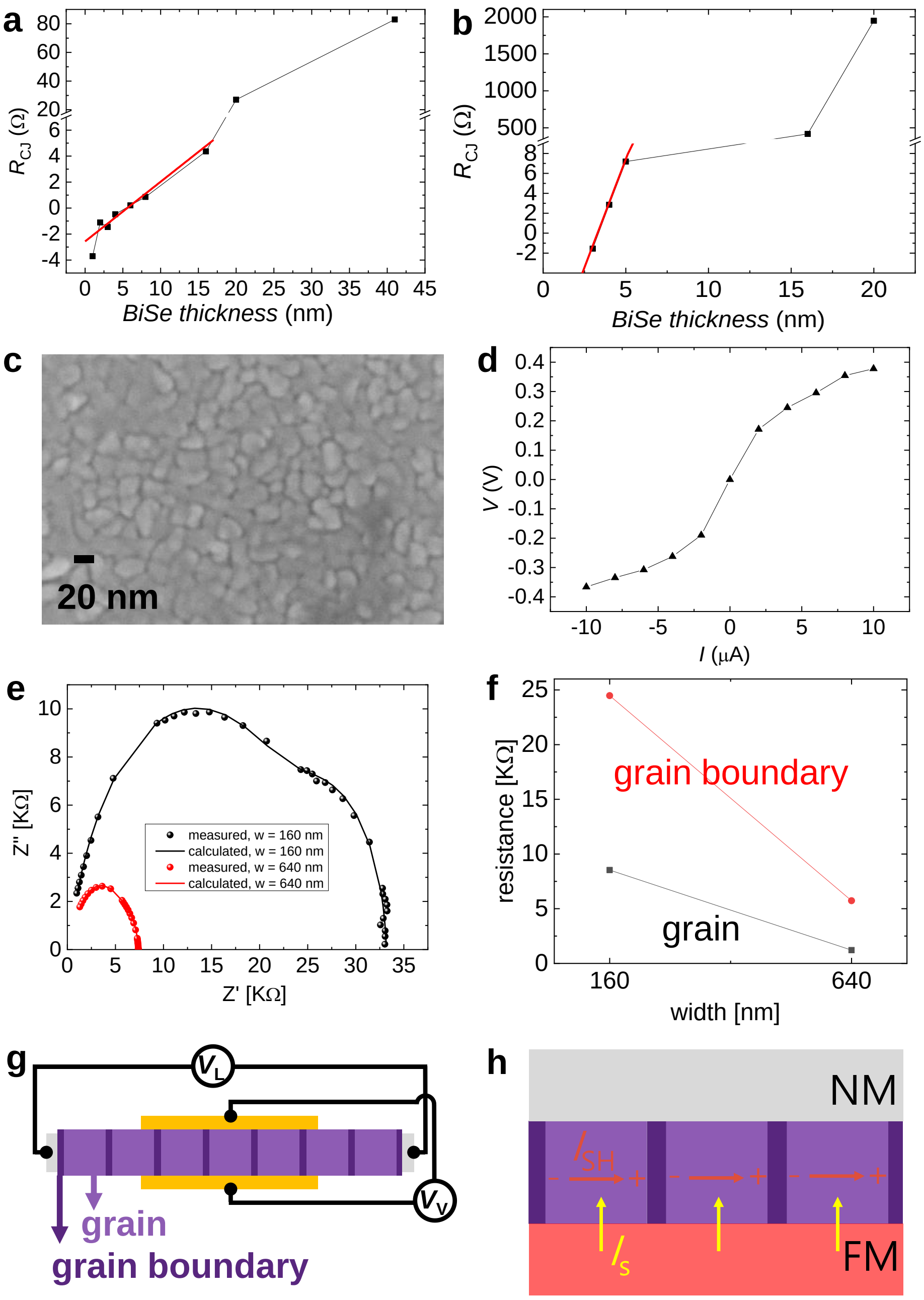

**Figure S7.** Grain and grain boundary of BiSe. Cross junction resistance as a function of BiSe thickness of (a) BiSe/Ti, and (b) BiSe/Pt structures. (c) SEM image of 32-nm-thick BiSe film. (d) Tunneling behavior in 32-nm-thick BiSe junction. (e) Nyquist plot of impedance analysis on BiSe wire and (f) two resistance terms as a fitting result. (g) Schematics of grain and grain boundary with measurement direction, lateral($V_L$) and vertical($V_V$). (h) Schematics of the SCC considering it occurs inside the grains.

Figure 3 of the main text shows that the resistivity of BiSe is changed by intermixing at the thickness of BiSe below 16 nm. On the other hand, even in the BiSe/Pt structure with limited intermixing, the resistivity of BiSe does not reach the 18,000 μΩ·cm obtained in a long nanowire (Figure 1f of the main text), which is thought to be due to the grain boundaries. Figure S7a and S7b show the $R_{CJ}$ of thicker BiSe in addition to the thickness range plotted in Figure 3a and 3b. The $R_{CJ}$ with BiSe thicker than 16 nm deviate significantly from the fitting line and increase more steeply, which means that BiSe resistivity increases.

Figure S7c is an SEM image of the grain structure of a 32-nm-thick BiSe film. The lateral grain size is 15-20 nm, and it is expected to have the same dimensions in the growth direction. Since BiSe has only one grain in the vertical direction below a thickness of 16 nm, $R_{CJ}$ is affected by the resistance inside the BiSe grain. However, above 16 nm, more than one grain is present, therefore boundaries which have higher resistivity are involved in the magnitude of $R_{CJ}$. The tunneling behavior shown in Figure S7d is sometimes observed in BiSe thickness over 16 nm, and it suggests the presence of a thin insulating layer inside. Note that applied current for Figure S7a and S7b is 10 μA in linear response.

For further verification, the impedance of a 16-nm-thick BiSe wire was measured by applying an alternating voltage of 30 mV in a frequency range from 1 kHz to 10 MHz and presented as a Nyquist plot in Figure S7e. In case of uniform conductor, only one arc appears, but as shown in Figure S7e, two arcs are partially overlapped, meaning that BiSe has two resistance components. Fitting was performed with EC-Lab software, and the resulting two resistance values are shown in Figure S6f. Considering the grain size of 15 nm and the thin grain boundary (< 2 nm), the resistance difference of about 2 times causes a big difference in resistivity.

Figure S7g shows why the resistivities measured in the BiSe wire and at the cross-junction are different. A resistance measurement in the wire is done in lateral direction ($V_L$), while $R_{CJ}$ is measured in vertical direction ($V_V$). In lateral (vertical) direction, grain and grain boundary exist as a series (parallel) resistance when BiSe is thinner than 16 nm, the maximum size of the grain. It explains the factor of 5, the difference between the BiSe wire (18,000 μΩ·cm) and the BiSe/Pt structure (3,700 μΩ·cm).

Since the grain boundary is much more resistive than the grain, most spin current will flow through the grain, and then will be spin-to-charge converted inside, as shown in Figure S6h. Induced voltage is within the grain, not at the grain boundary, which means that even though the obtained resistivity is high due to grain boundary, the signal magnitude of SCC is dependent on the resistivity of the grain.

As shown in Figure S7a and S7b, the BiSe resistivity increases due to the grain boundary at thicknesses above 16 nm. However, since it is also affected by intermixing, the resistivity varies depending on NM. In the case of the BiSe/Pt structure, the resistivity recovers to that of BiSe wire when it is calculated using the $R_{CJ}$ value, thickness, and junction area. Also, since the grain boundary acts as a shunting barrier, it interferes with the observation of the SCC signal in the local spin injection device.

## Supporting References